\documentclass[aps,prb,twocolumn,groupedaddress,showpacs]{revtex4}
\usepackage{graphicx}
\begin{document}


\title{Electronic structure, magnetism and spin fluctuations in the superconducting weak ferromagnet Y$_{4}$Co$_{3}$}


\author{B. Wiendlocha}
\email[corresponding author: ]{bartekw@fatcat.ftj.agh.edu.pl}
\author{J. Tobola}
\author{S. Kaprzyk}
\author{A. Kolodziejczyk}
\affiliation{Faculty of Physics and Applied Computer Science, AGH University of Science and 
Technology, Al. Mickiewicza 30, 30-059 Cracow, Poland}


\date{\today}

\begin{abstract}
Results of the first principles study on the electronic structure and magnetism of the superconducting 
weak ferromagnet Y$_4$Co$_3$, are presented. Using the full potential Korringa-Kohn-Rostoker 
(FP-KKR) method, densities of states, dispersion curves and magnetic moments were calculated for 
quasi-ordered structural model of the compound in the framework of the local density approximation. 
Spin-polarized KKR calculations confirm that weak ferromagnetic properties of Y$_4$Co$_3$ can 
be attributed to only one cobalt atom located on (2b) site in the unit cell, while other twenty Co and Y atoms acts as a diamagnetic environment. 
Moreover, the magnetic Co atoms form a quasi-one-dimensional chains along $z$ direction.
The magnitude of Co(2b) magnetic moment ($0.55~\mu_B$) markedly overestimates the experimental 
value (0.23~$\mu_B$), which suggests the importance of spin fluctuations in this system. 
Calculated distribution of spin magnetization in the unit cell provides a background for discussion 
of the coexistence of ferromagnetism and superconductivity in Y$_4$Co$_3$. Finally, the effect 
of pressure on magnetism is discussed and compared with experimental data, also supporting 
weak ferromagnetic behaviors in the system.
\end{abstract}


\maketitle




\section{Introduction}
Unusual properties of Y$_4$Co$_3$, that is the coexistence of superconductivity and ferromagnetism, were observed 30 
years ago.~\cite{akolo-80} The superconducting and ferromagnetic critical temperatures are about $T_s \sim 2.5$~K and $T_C \sim 4.5$~K, respectively.
After this finding, the system was intensively studied using different techniques, since it was the first 
example where superconductivity coexisted with weak itinerant ferromagnetism, with both phenomena driven most likely by the $d$-band electrons. 
To our best knowledge, Y$_4$Co$_3$ is up to now a unique compound containing transition-metal elements only, where coexistence 
of ferromagnetism and superconductivity occurs, since other systems always contain $f$-like elements, e.g. 
UGe$_2$ (Ref.~\onlinecite{uge2}), URhGe (Ref.~\onlinecite{urhge}), UCoGe (Ref.~\onlinecite{ucoge}) or recently discovered P-doped EuFe$_2$As$_2$ (Ref.~\onlinecite{eufe2as2-p}).
Moreover, Y$_4$Co$_3$ was likely the first example, where superconductivity occurred below a transition to 
ferromagnetism, with both phenomena coexisted at least in the range between 1~K and 2.5~K.~\cite{yco-akolo,yco-acta} 
This was another difference to previously known magnetic superconductors e.g. ErRh$_4$B$_4$ or HoMo$_6$S$_8$ 
(see, Refs.~\onlinecite{errh4b4,homo6s8-1,homo6s8-2}), where magnetism appeared below $T_s$, 
suppressing superconductivity.

\begin{figure}[b!]
\includegraphics[width=0.48\textwidth]{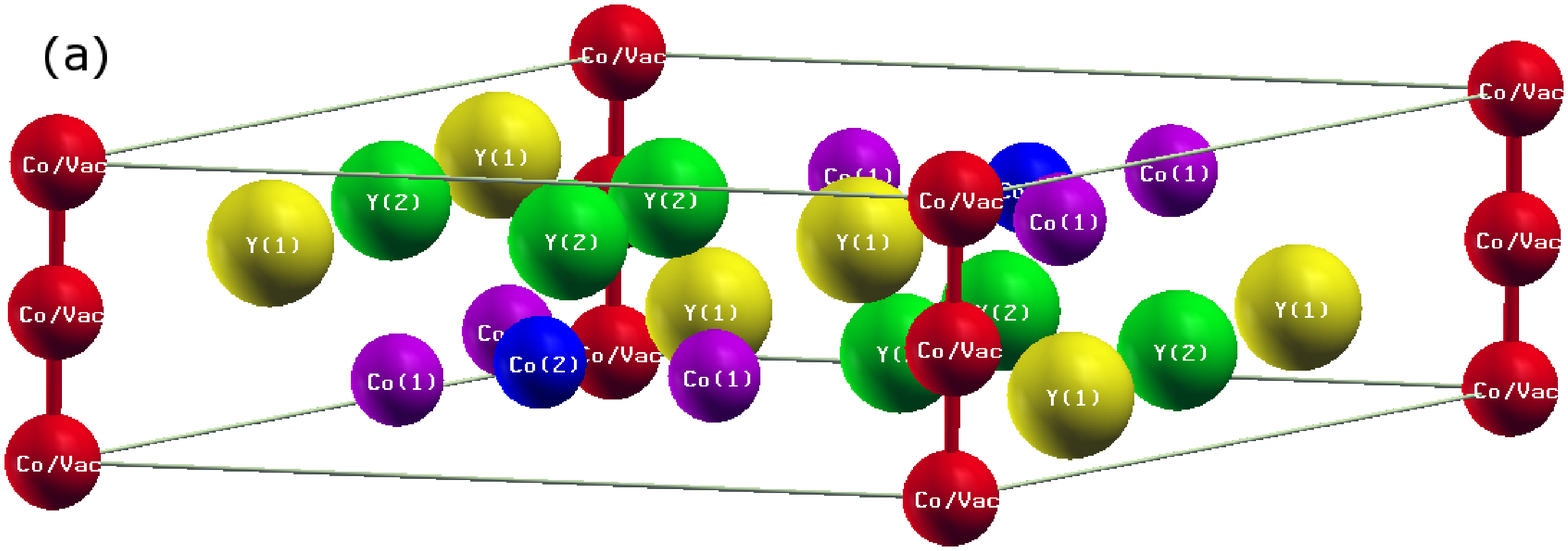}\\
\includegraphics[width=0.34\textwidth]{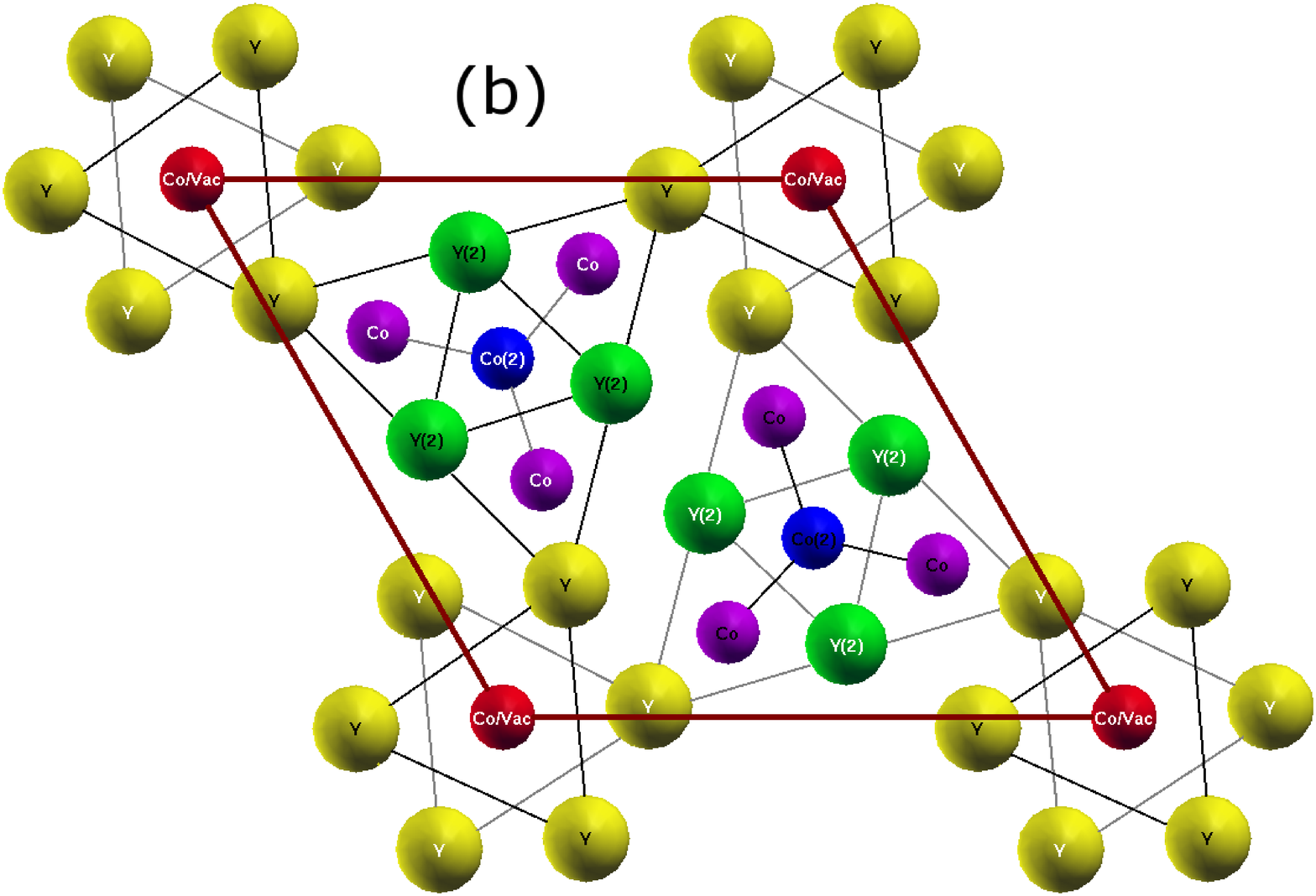}\raisebox{12pt}{\includegraphics[width=0.14\textwidth]{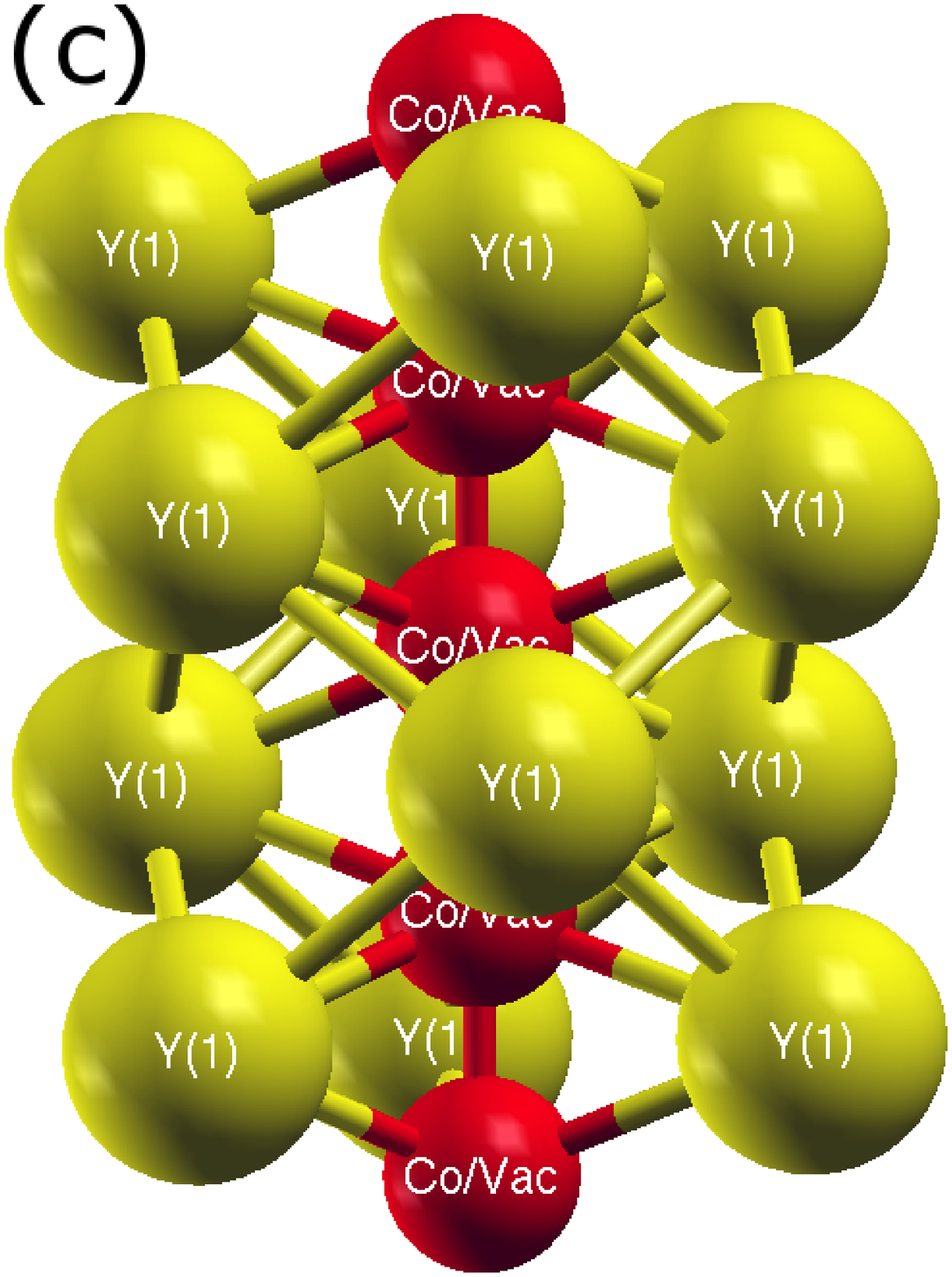}}
\caption{(Color online) (a) Unit cell of Y$_4$Co$_3$ used in the calculations. The edges of the unit cell, i.e. (2b) sites, are occupied by cobalt or vacancy (in red), on average every second position is occupied by cobalt, called here Co(3). Atomic positions are presented in Table~\ref{yco-pos}. 
(b) The body of the unit cell is formed by the trigonal prisms of Y(1) (yellow), Y(2) (green), Co(1) (violet) and Co(2) (blue), the black lines and labels denotes atoms in the $z=0.75$ plane, while white denotes $z = 0.25$ plane. (c) The in-plane triangles of Y(1) atoms form channels along $z$ direction, surrounding the Co(3) atoms chains.}\label{fig:struct}
\end{figure}

Although Y$_4$Co$_3$ has been known since 1980, the first principles study on its magnetism (e.g. calculations of 
magnetic moments) has not been presented until now. We have found  two papers reporting electronic 
structure of this system from {{\it ab initio}} calculations. Interesting discussion of electronic structure of related compound Y$_9$Co$_7$ based on non-spin-polarized computations can be found in Ref.~\onlinecite{yco-shimizu} (for connections between Y$_4$Co$_3$ and Y$_9$Co$_7$, see below). More recent paper, Ref.~\onlinecite{yco-jeong}, briefly reports electronic structure calculations for Y$_4$Co$_3$, also in non-magnetic state. 

Noteworthy, one may notice a renewed interest on this system due to very recent experimental 
investigations.~\cite{klimczuk-conf}

In this paper the results of band structure calculations for Y$_4$Co$_3$ are presented. 
We attempt to answer some open questions, e.g. whether the magnetism is an intrinsic property of the system or an effect 
of crystal imperfections,~\cite{yco-shimizu} or what is the microscopic reason for magnetism occurrence in Y$_4$Co$_3$. 
The implications of our electronic structure calculations on the current model of the coexistence of ferromagnetism and 
superconductivity in Y$_4$Co$_3$ are also discussed.

\subsection{Review of the experimental data}

In this section we summarize shortly crystal structure and magnetism data, available for Y$_4$Co$_3$.
For a more complete review, see e.g. Refs.~\onlinecite{yco-akolo,yco-acta}. The crystal structure of the Y-Co system near 1:1.3 
stoichiometry is rather complex, and was oryginally described in the hexagonal unit cell of the Ho$_4$Co$_3$ type~\cite{yco-yvon} (Fig.~\ref{fig:struct}). 
Unit cell can take 22 atoms, being distributed over three inequivalent Co sites and two inequivalent Y sites. All positions are 
gathered in Table~\ref{yco-pos}. As the unit cell includes three formula units, for the Y:Co stoichiometry equals 
to 4:3, the Co(2b) sites are half-filled ($50\%$) and the number of atoms in the unit cell is equal to 21. 
Thus, in this crystallographic model, Y$_4$Co$_3$ can not be 
regarded as an ordered compound, but as an disordered alloy with (2b) sites occupied randomly by cobalt (Co atom on (2b) site will be called Co(3)) and 
vacancies (Vac). Thus, the crystal cell of this alloy better corresponds to the formula Y$_{12}$Co$_{8+2x}$ 
with $x = 0.50$ than to the formula Y$_4$Co$_3$, suggesting an ordered system. 
The unit cell (Fig.~\ref{fig:struct}) consists of the 'separated' Co/Vac chains forming the cell edges and 
the Y(1)-Y(2)-Co(1)-Co(2) trigonal prisms filling most of the unit cell. The partial occupation of 
the Co(3) site is likely connected with a short distance between the neighboring (2b) positions ($\sim 2~$\AA). 

In 80-ties, after revealing the coexistence of ferromagnetism and superconductivity in Y$_4$Co$_3$, the Ho$_4$Co$_3$-type 
unit cell appeared to be only an approximation for the real structure of the system. There was a debate whether 
this system should be called Y$_9$Co$_7$ (instead of Y$_4$Co$_3$), since it was found,~\cite{yco-akolo,yco-acta,yco-big} 
that Y$_9$Co$_7$ (richer in cobalt) exhibited better superconducting properties ($\sim 0.5$~K higher $T_s$) 
than Y$_4$Co$_3$. It was even suggested, that Y$_4$Co$_3$ as a single phase compound might not exist.~\cite{yco-big} 
Employing the initial notation, the 9:7 stoichiometry compound corresponds to Y$_4$Co$_{3.11}$, but one should bear 
in mind that for Y$_9$Co$_7$ an ordered model of structure was suggested~\cite{yco-big} with the unit 
cell tripled along the $c$-axis. In such a superstructure, the Co(3) atoms positions are adapted in such way to make the Y-Co system perfectly ordered for 9:7 composition (100\% filling of all sites, see e.g. Refs.~\onlinecite{yco-shimizu,yco-big} for details). 
The resulting crystal structure exhibits reduced symmetry, trigonal instead of hexagonal.~\cite{yco-yvon} 
However, the ordered model of the superstructure was not confirmed by X-ray measurements on single crystal 
samples.~\cite{yco-yvon} Moreover, it was also reported~\cite{yco-yvon} that the Co(3) atoms did not occupy the 
fixed lattice positions, and were rather arranged in chains with undetermined period. Thus, the basic unit cell shown in 
Fig.~\ref{fig:struct} can be treated as the best established and simplest approximation of the real structure both for Y$_4$Co$_3$ and Y$_9$Co$_7$, as far as the Co(3) atoms location is concerned. 
The Y$_{12}$Co$_{8+2x}$ model represents the Y$_4$Co$_3$ and Y$_9$Co$_7$ compounds, when the Co(3) concentration reaches the values of $x = 0.50$ and $x = 0.66$, respectively. However, the question on the real unit cell of Y$_4$Co$_3$ and 
Y$_9$Co$_7$ and whether they describe the same crystal structure with various Co contents, is still open. 

In the present work, the basic unit cell, presented in Fig.~\ref{fig:struct}, with $x = 0.50$ was employed to account for 
the Y$_4$Co$_3$ case. This crystal model has allowed us to use the full potential Korringa-Kohn-Rostoker 
code adapted to electronic structure calculations of ordered compounds (see, Section~\ref{sec.calc}).

\begin{table}[t]
\caption{Atomic positions of the Y$_4$Co$_3$ compound (type Ho$_4$Co$_3$, space group No. 176, $P63/m$). 
Lattice parameters are:~\cite{yco-szytula} $a$ = 11.527~\AA, $c$ = 4.052~\AA.} \label{yco-pos}
\begin{ruledtabular}
\begin{tabular}{lcccc}
Atom & site & \multicolumn{1}{c}{coordinates} & $x$ & $y$ \\
\hline
Y(1) & (6h) & $(x,y,{1\over4})$, $(\bar{y},x$-$y,{1\over4})$ $(y$-$x,\bar{x},{1\over4})$ & 0.7543 & 0.9791   \\
&&$(\bar{x},\bar{y},{3\over4})$ $(y,y$-$x,{3\over4})$, $(x$-$y,x,{3\over4})$ && \\ 
Y(2) & (6h) & as above & 0.1360 & 0.5150  \\
Co(1) & (6h) & as above  & 0.4415 & 0.1578  \\
Co(2) & (2d) & $({2\over 3}, {1\over3}, {1 \over 4})$, $({1 \over 3}, {2\over3},{3\over4})$ & -- & -- \\
Co(3) & (2b)\footnotemark[1]  & $(0,0,0)$, $(0,0,{1 \over2})$ &--& -- \\
\end{tabular}
\end{ruledtabular}
\footnotetext[1]{In our calculations (2b) positions are split into (1a) $(0,0,0)$ and (1b) $(0,0,{1 \over2})$, being 
occupied by Co and vacancy, respectively (see text).}
\end{table}

As far as the superconducting properties are concerned, it is generally believed, that 
superconductivity in Y$_4$Co$_3$ has a singlet-like character and probably mediated by the 
electron-phonon interactions.~\cite{yco-akolo} The system exhibits all the characteristic features of 
conventional superconductors (see Refs.~\onlinecite{yco-akolo,yco-acta} and references therein), 
e.g. perfect diamagnetic Meissner state, the specific heat jump at $T_s$ correlated with drop 
in resistivity and susceptibility.
More puzzling seems the type of its magnetism. The system shows e.g. a modified Curie-Weiss behavior, 
with susceptibility maximum ~\cite{yco-spalek} and specific heat jump near the 
Curie temperature,~\cite{yco-tarnawski} suppression of magnetism with pressure~\cite{yco-press} 
and a hysteresis loop in the magnetization measurements~\cite{yco-low_field} interpreted as 
fingerprints of a ferromagnetic state. 
There are different values of magnetization reported in literature. For the 4:3 composition one can find 
$M = 0.045$~$\mu_B$/f.u. measured for Y$_4$Co$_{3.03}$ sample~\cite{yco-ni} and extrapolated 
to $T=0$K or $M \simeq 0.03$~$\mu_B$/(f.u. Y$_4$Co$_3$) at 1.24~K.~\cite{yco-low_field} Note, that these 
values are not exact, firstly diamagnetism induced by superconductivity below 2.5~K does not allow to 
accurately measure bulk magnetization, and secondly it is difficult to measure magnetization in 
weak ferromagnetic system.~\cite{yco-ni}

From the NMR measurements~\cite{yco-nmr1}  it was found that there were three inequivalent Co sites in Y$_4$Co$_3$, but 
magnetism was attributed only to the cobalt atoms on (2b) site. The estimated magnetic moment was
$\mu_{Co(3)} = 0.23$~$\mu_B$, and important spin fluctuations were observed. 
Actually, there are no other experimental papers on determination of the Co(3) magnetic moment, thus our 
theoretical result is compared to this value. Most of the experimental results concerning magnetism of 
the Y-Co system were interpreted in the framework of the spin-fluctuation theory of the weak itinerant ferromagnetism,~\cite{moriya-book} showing characteristic features as e.g. the zero-field magnetization $M(T)^2 \propto T^{4/3}$ (Ref.~\onlinecite{yco-ni}) or resistivity $\rho \propto T^2$ (Ref.~\onlinecite{yco-spalek}).
However, more exotic models without long-range ferromagnetic ordering were also considered, i.e. 
basing on the $\mu$SR measurements~\cite{yco-usr} the magnetic state of Y$_9$Co$_7$ was called 
'crypto-itinerant ferromagnet', some hybrid magnetic and superconducting state was 
suggested,~\cite{yco-hybrid} while short range order was discussed elsewhere.~\cite{yco-sro}

On the whole, magnetic properties of the Y-Co system near 4:3 or 9:7 stoichiometry were found to be similar 
and weakly dependent on Y:Co composition~\cite{yco-liet} (comparing to superconducting behaviors), maybe except 
for increasing magnetization with increasing Co content.~\footnote{Magnetization measured for 
Y$_9$Co$_7$ sample and extrapolated to T = 0~K was $M \simeq 0.383$~emu/g, which referred to 
the 4:3 composition gives about  $M= 0.10$~$\mu_B$/(f.u. Y$_4$Co$_3$, and is higher than in Y$_4$Co$_3$ likely due to higher 
Co(3) atoms concentration.~\cite{yco-sulkowski}} 
All these data allow to suppose that general conclusions deduced from the first principles calculations 
for Y$_4$Co$_3$ should also be maintained for Y$_9$Co$_7$ (e.g. we may expect similar local magnetic 
moments). Also, some experimental results available only for Y$_9$Co$_7$ (e.g. effect of pressure on magnetism), 
should provide reliable comparison with theoretical results obtained for the quasi-ordered model 
of Y$_4$Co$_3$.

To summarize this short review, the commonly accepted phenomenological model of the coexistence of 
superconductivity and magnetism in Y$_4$Co$_3$ / Y$_9$Co$_7$ system is as follows: there is an 
itinerant weak ferromagnetic state below $T_C$ with the Co(3) sublattice responsible for magnetism, combined with 
superconductivity below $T_s$, being mostly attributed to the Y-Co triangular prisms inside the unit cell. 
Both physical phenomena visibly compete (e.g. external pressure suppresses magnetism and enhance 
superconductivity~\cite{yco-press,yco-press-apl}), but their coexistence is possible due to some spatial 'separation' of atom sublattices responsible for different phenomena.~\cite{yco-akolo,yco-acta} However, this separation cannot be treated in the strict sense, since the Ginzburg-Landau coherence length was estimated to 
be as large as $300~\AA$ (roughly thirty times larger than the unit cell size).~\cite{akolo-coherence} 
One can tentatively imagine this uncommon state as a 'superconducting sea' with the Co(3) 'magnetic islands' embedded. 
The question on the existence of the 'magnetic islands' and whether Co(3) moments are ordered in Y$_4$Co$_3$ 
has not been yet addressed to first principles calculations. Hence, the KKR results presented here may help to 
enlighten this problem.

\section{Computational details}\label{sec.calc}

Electronic structure calculations were performed using the {\it full potential} Korringa-Kohn-Rostoker (FP-KKR) 
technique based on the Green function multiple scattering theory.~\cite{kkr99, stopa, phd} In our implementation 
of the FP-KKR method, the unit cell is divided into the set of generalized Voronoi polyhedrons arbitrary formed around
inequivalent sites, that completely fill the Wigner-Seitz cell. The crystal potential has been constructed in the framework of the 
local spin density approximation (LSDA), using Perdew-Wang formula~\cite{pw} for the exchange-correlation part. 
All results were carefully checked for the \makebox{{\bf k}-point} number convergence, using more than 100 points 
in the irreducible part of Brillouin zone for the tetrahedron integration method. The results of FP-KKR semi-relativistic calculations are presented, with the 
fully relativistic treatment of core electron energy levels.

\begin{figure}[b]
\includegraphics[width=0.48\textwidth]{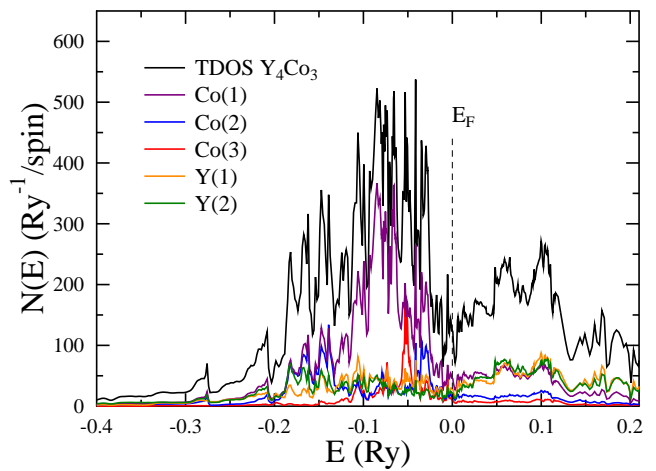}
\caption{\label{fig-tdos1}(Color online) Total and site-decomposed densities of electronic states in Y$_4$Co$_3$ (per unit cell, i.e. three formula units), from non-spin-polarized calculations.}
\end{figure}

In this work, the ordered model of the hexagonal cell (Fig.~\ref{fig:struct}) was used in the electronic structure 
calculations. To take into account the partial filling of the Co(2b) site in Y$_4$Co$_3$ in KKR method, we employed
a 'quasi-disordered' structure, i.e the (2b) position was split into two non-equivalent sites, (1a) and (1b). 
The (1a), i.e. $(0,0,0)$ position, was occupied by Co atom (Co(3)) and (1b), $(0,0,{1 \over 2})$ site, by a vacancy (Vac), i.e. 
an empty sphere with $Z=0$. This structural modification resulted in the lowering of the hexagonal cell symmetry 
from space group  $P63/m$ (No. 173) to $P$-3 (No. 147), with reduction of symmetry operations from 12 to 6. 
As above-mentioned, this model admits to consider the Y$_4$Co$_3$ composition only. 
It was verified by additional calculations that the KKR results remain unchanged when the positions of Co(3) and 
Vac were exchanged (cobalt on (1b) site and  empty spheres on (1a) site), as one expects from equivalence 
of both sites in the proper space group.

\section{RESULTS AND DISCUSSION}

\subsection{Non spin-polarized electronic structure}

\begin{figure}[b]
\includegraphics[width=0.49\textwidth]{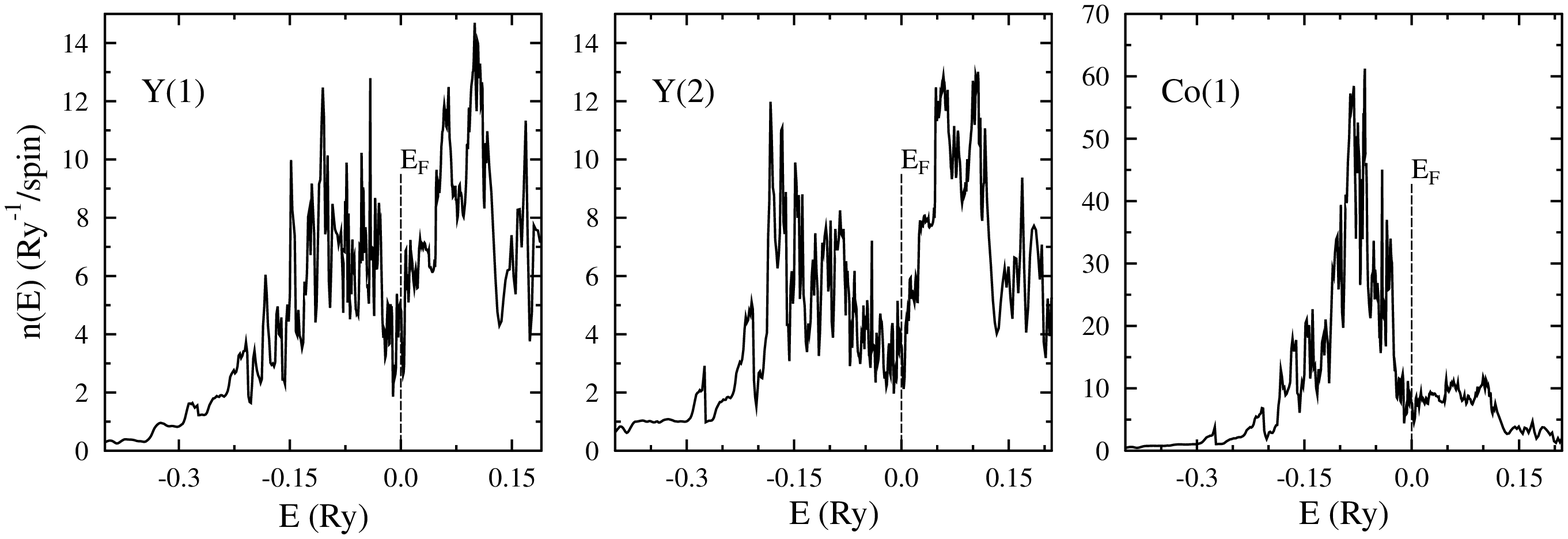}
\includegraphics[width=0.49\textwidth]{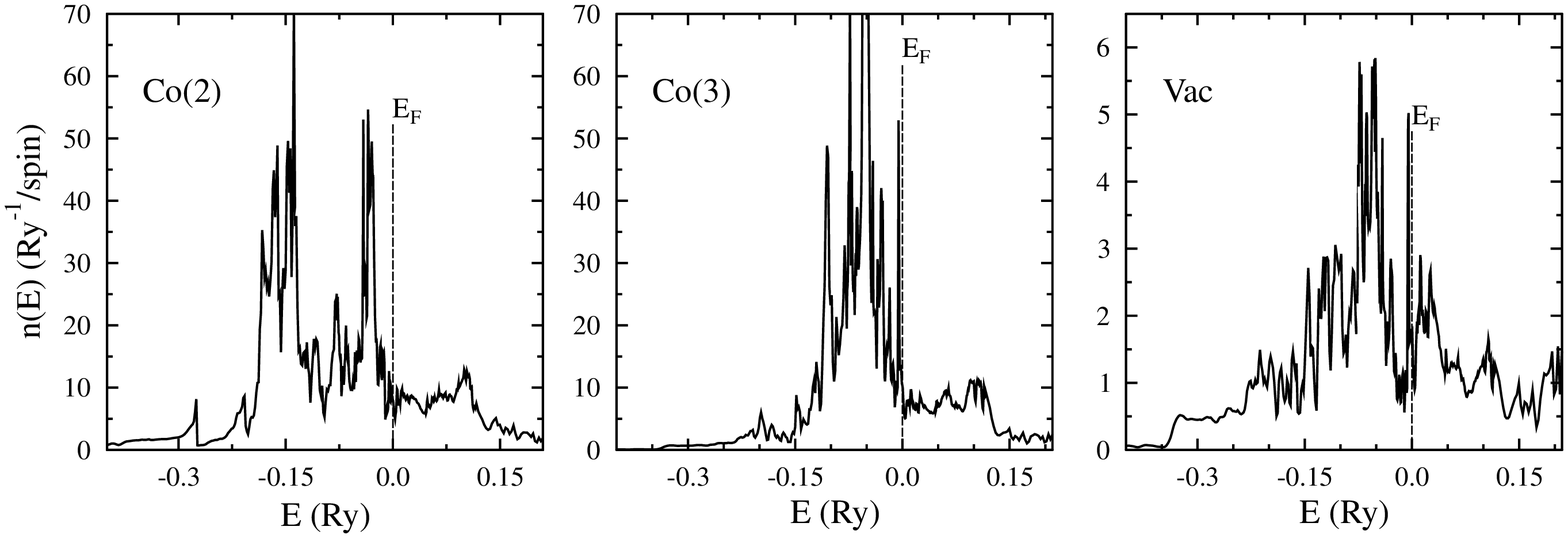}
\caption{\label{fig-pdos}Site decomposed DOS per atom from non-spin-polarized calculations.}
\end{figure}

\begin{figure}[b]
\makebox{\includegraphics[width=0.24\textwidth]{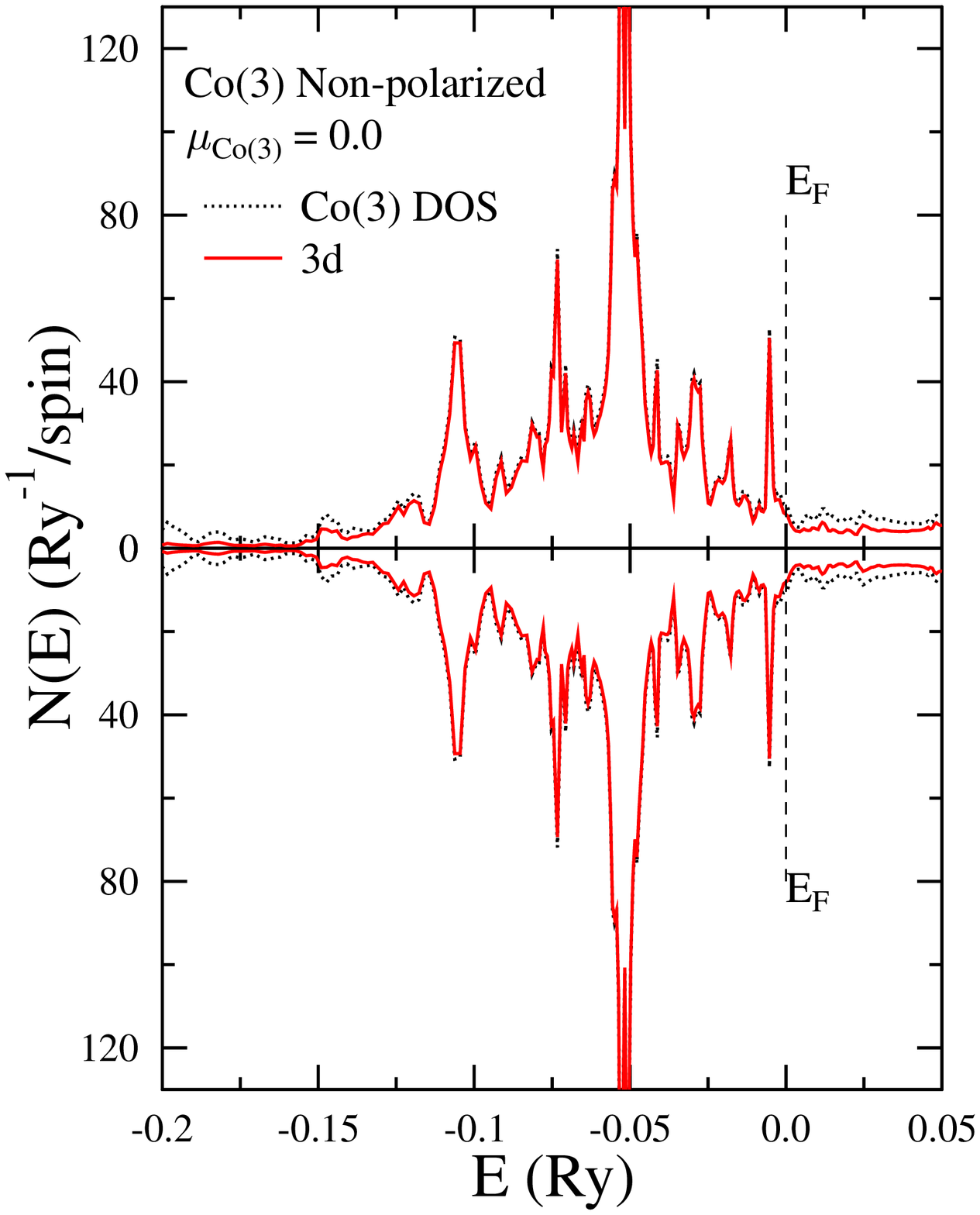} 
\includegraphics[width=0.24\textwidth]{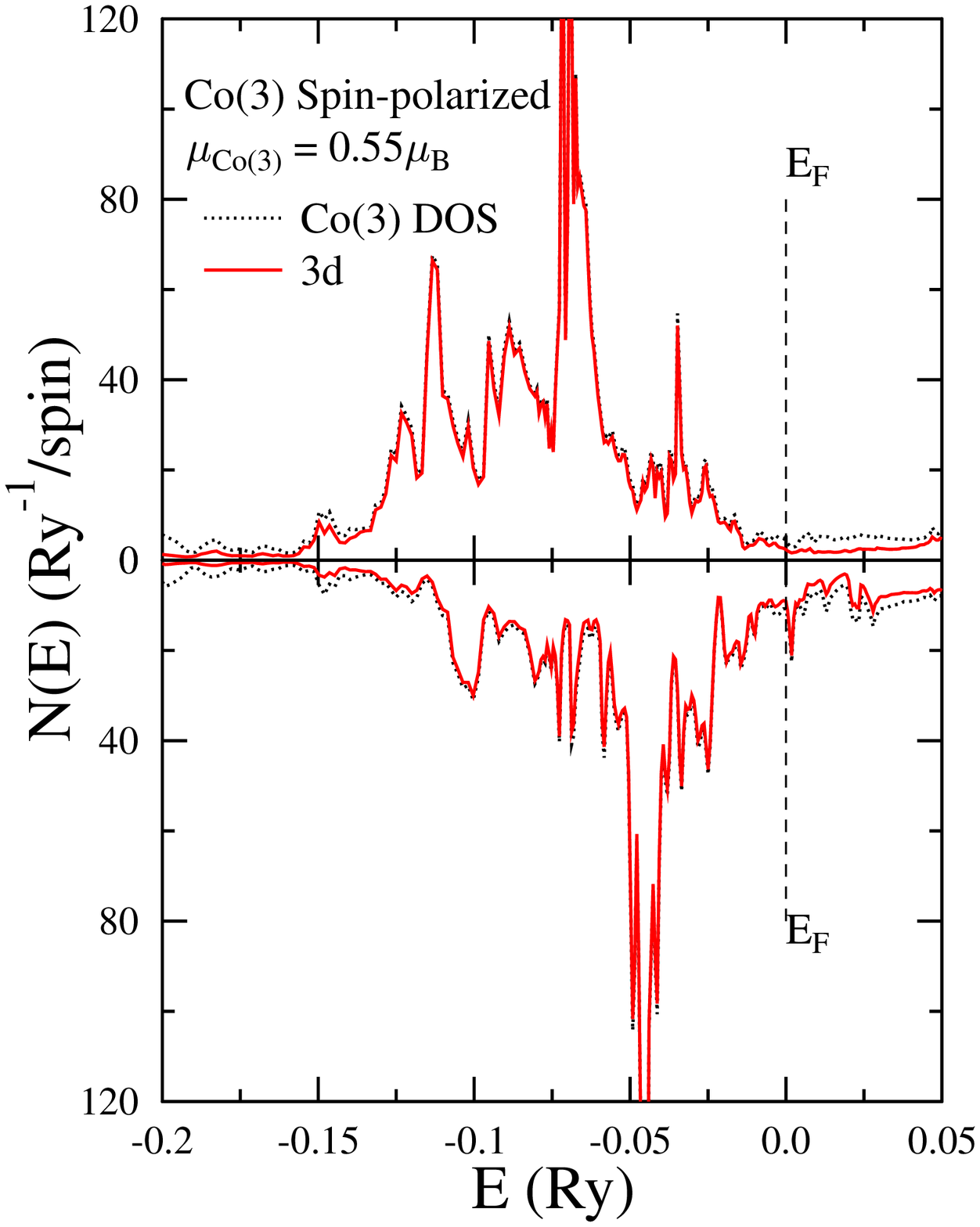}}
\caption{(Color online) Comparison of the Co(3) partial DOS from non-spin-polarized (left) and spin-polarized (right) calculations. Peak under $E_F$ in non-magnetic state is well visible.}\label{yco-co3-dos}
\end{figure}

The discussion of electronic structure of Y$_4$Co$_3$ is started from analysis of general features in non spin-polarized case. 
The total and site-decomposed electronic densities of states (DOS) are presented in Fig.~\ref{fig-tdos1}. 
Due to the large number of atoms in the unit cell and in consequence various interatomic distances, the DOS has a  
complex shape. The Fermi level ($E_F$) is located in the DOS valley, which can tentatively support 
the chemical stability of this system. The valence band region is formed from the strongly hybridized atomic states of 
Co ($3d$, $4s$) and Y ($4d$, $5s$), with important $s-p$ and $s-d$ charge transfers. 
The Y-$4p$ states constitute semi-core level, located about -1.5~Ry below the Fermi level (not shown). 
Generally, the body of the electronic structure is dominated by contributions from the Y(1)-Y(2)-Co(1)-Co(2) block, 
which form the triangular prisms inside the unit cell (Fig.~\ref{fig:struct}). Nominally, the major contribution to DOS comes 
from six Co(1) atoms, due to the highest number of valence electrons (54 $e$) given to the bands. 
The single Co(3) atom, building the separated cobalt chains, roughly governs the position of the Fermi 
level due to DOS significantly different from other atoms (a high $d$-like peak below $E_F$), what enables 
the Co(3) to play important role in ground state properties of Y$_4$Co$_3$.

\begin{table}[t]
\caption{Electronic properties of Y$_4$Co$_3$. Values of DOS $n(E_F)$ are given in (Ry$^{-1}$/spin), magnetic moments $\mu$ in ($\mu_B$), $S$ stands for the Stoner parameter.} \label{tab:dos}
\begin{ruledtabular}
\begin{tabular}{lrrrrc}
\multicolumn{6}{c}{non--spin--polarized calculations} \\ \hline
Atom & \multicolumn{2}{c}{$n(E_F)$} & \multicolumn{2}{c}{$n_{d}(E_F)$}  &  $S$ \\
\hline
Y(1)  &  \multicolumn{2}{c}{3.62} & \multicolumn{2}{c}{2.50}  & 0.15 \\
Y(2)  &  \multicolumn{2}{c}{4.85} & \multicolumn{2}{c}{3.37}  & 0.19 \\
Co(1) &  \multicolumn{2}{c}{7.71} & \multicolumn{2}{c}{5.79}  & 0.43 \\
Co(2) &  \multicolumn{2}{c}{8.00} & \multicolumn{2}{c}{6.86}  & 0.48 \\ 
Co(3) &  \multicolumn{2}{c}{10.31} & \multicolumn{2}{c}{8.19} & 0.61 \\ 
Vac   &  \multicolumn{2}{c}{1.73} & \multicolumn{2}{c}{0.47}  & --- \\ 
\multicolumn{6}{c}{$N(E_F) = 250$~Ry$^{-1}$ per unit cell} \\ \hline \hline
\multicolumn{6}{c}{spin--polarized calculations} \\ \hline
Atom & $n_{\uparrow}(E_F)$ & $n_{\downarrow}(E_F)$ & $n_{\uparrow d}(E_F)$ & $n_{\downarrow d}(E_F)$ & $\mu$ \\ \hline
Y(1)   &  4.00 &  4.22 & 2.72 & 2.91 & -0.024 \\
Y(2)   &  2.90 &  4.02 & 1.96 & 2.92 &  0.004  \\ 
Co(1)  & 6.37 &  7.18 & 4.83 & 5.18 & -0.007 \\
Co(2)  &  6.45 &  7.38 & 5.49 & 6.41  & -0.017 \\
Co(3)  &  4.26 &  10.75 & 2.37 & 9.14 &  0.551  \\ 
Vac    & 1.07 &  1.15 & 0.39 & 0.37   &  0.016 \\ 
\multicolumn{6}{c}{$N_{\uparrow}(E_F) = 99$~Ry$^{-1}$ $N_{\downarrow}(E_F) = 120$~Ry$^{-1}$ }\\
\end{tabular}
\end{ruledtabular}
\end{table}

To have a deeper insight into influence of each atom on electronic structure, the site-decomposed DOS 
are plotted for inequivalent sites (Fig.~\ref{fig-pdos} and Fig.~\ref{yco-co3-dos}). The most striking DOS feature of Y$_4$Co$_3$ 
is that except for Co(3), the Fermi level is systematically placed in the local DOS valleys, resulting in 
relatively low total DOS. The calculated small DOSs at $E_F$ per atom (see, Table~\ref{tab:dos}) 
would rather not suggest a transition to ferromagnetic state. In contrast to other atoms, Co(3) exhibits 
apparently different DOS, due to a large $d$-like peak found just below the Fermi level ($E_F$ is placed on 
the decreasing DOS slope). This feature yields the highest DOS value (per atom) for Co(3) in comparison to
other contributions (Table~\ref{tab:dos}). Actually, the value $n(E_F) \simeq 10$~Ry$^{-1}$/spin is too 
small to satisfy the Stoner criterion, since the calculated Stoner parameter for Co(3) atom $S = I_d n_d(E_F)$ is 
only about 0.6. Thus, the Stoner analysis based on non-spin-polarized DOS and the exchange integral 
predicts a non-magnetic ground state of Y$_4$Co$_3$. However, the presence of large and narrow DOS peak in the 
vicinity of $E_F$ causes that the simple Stoner criterion of ferromagnetism onset should be taken with care and 
accurate spin-polarized calculations may determine the preferred ground state.

\subsection{Ferromagnetism}

\begin{figure}[t]
\includegraphics[width=0.48\textwidth]{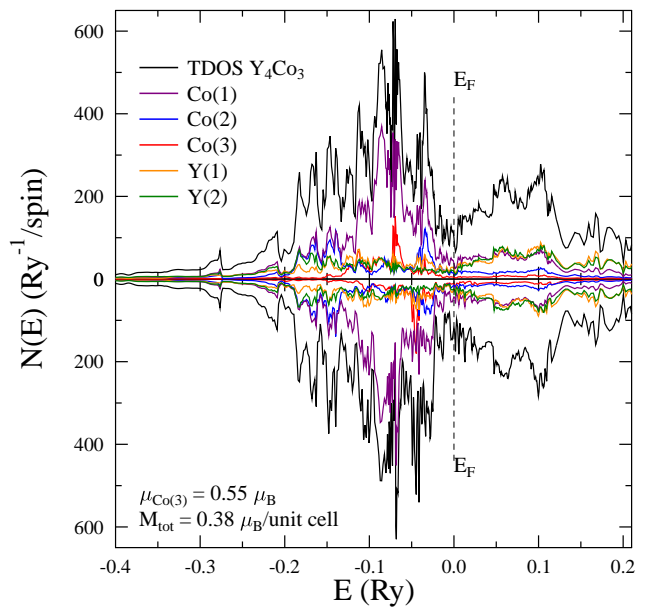}
\caption{\label{fig-tdos2}(Color online) Total and site-decomposed densities of electronic states in Y$_4$Co$_3$  from spin-polarized calculations.}
\end{figure}

Indeed, the spin-polarized calculations do not confirm the non-magnetic state deduced from the Stoner criterion. 
Figure~\ref{fig-tdos2} presents the spin-polarized DOS shape of Y$_4$Co$_3$. At first glance, the differences between the non 
spin-polarized (Fig.~\ref{fig-tdos1}) and spin-polarized (Fig.~\ref{fig-tdos2}) total DOSs are hardly visible, 
since spin-up and spin-down DOS functions are very similar. However, the spin-polarized KKR calculations finally 
resulted in stable, ferromagnetic ground state. The partial, site-decomposed DOSs (Fig.~\ref{fig-tdos2}) show that 
only the Co(3) atom exhibits polarization in DOS shape, as depicted in Fig.~\ref{yco-co3-dos}, being 
confirmed by appearance of local magnetic moment (Table~\ref{tab:dos}). The Co(3) magnetic moment  
of about $\mu_{Co_3} = 0.55$~$\mu_B$ decides in favor of magnetism in Y$_4$Co$_3$, since the remaining atoms possess small
magnetic moments ($\leq 0.02~\mu_B$). Such a small DOS polarizations seen on other atoms should be 
rather considered as a response to magnetic field of Co(3) atom, than an intrinsic local magnetic moments.
The total magnetization $M_{tot} = 0.38$~$\mu_B$/unit cell yields $M \simeq 0.13$~$\mu_B$ 
per Y$_4$Co$_3$ formula unit. $M_{tot}$ is lower than $\mu_{Co_3}$ due to the overall 'diamagnetic' response of 
the Y-Co triangular prisms, generating about -0.1 $\mu_B$. The comparison with the experimental values is 
discussed in Section~\ref{spinfl}.

\begin{figure*}[htb]
\includegraphics[width=0.38\textwidth]{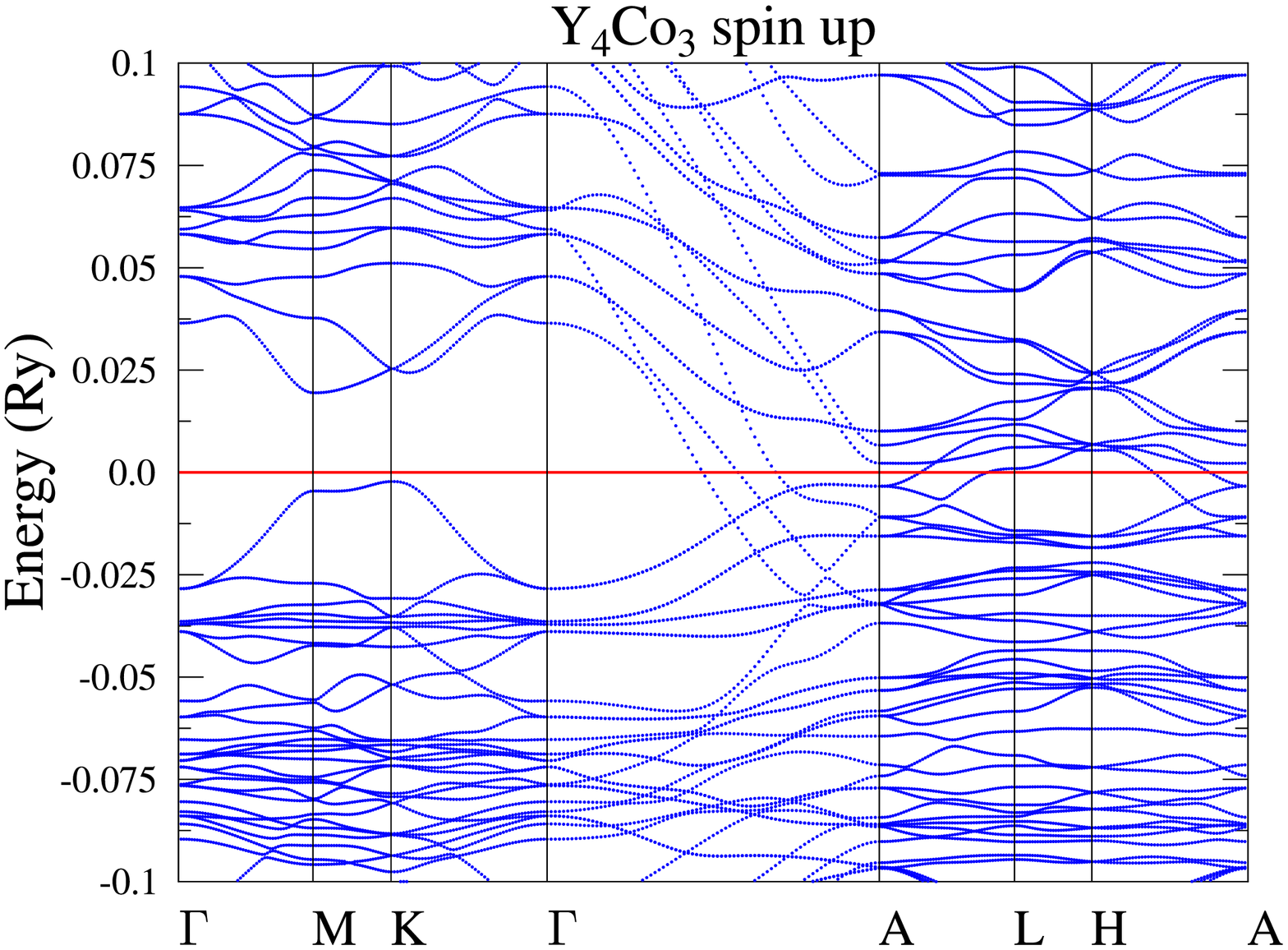}
\raisebox{100pt}{\includegraphics[width=0.15\textwidth,angle=-90]{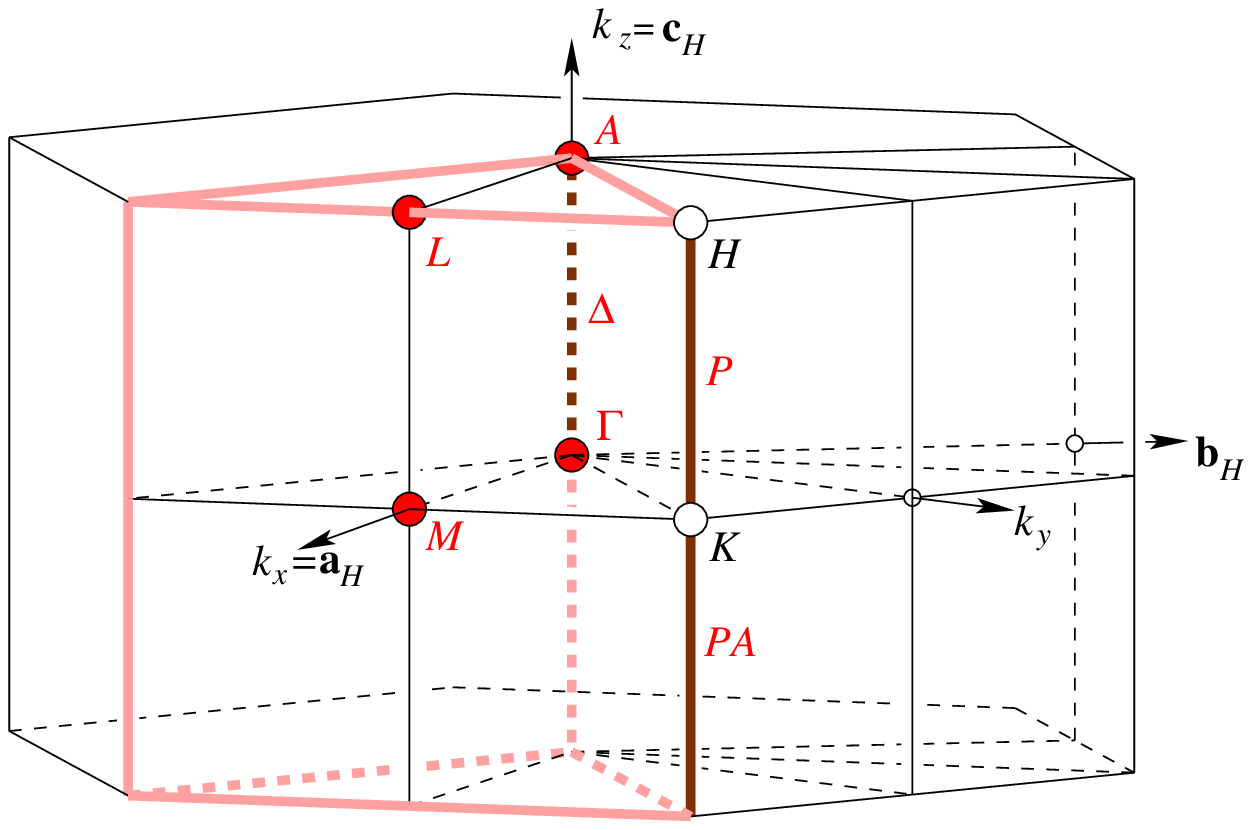}}
\includegraphics[width=0.38\textwidth]{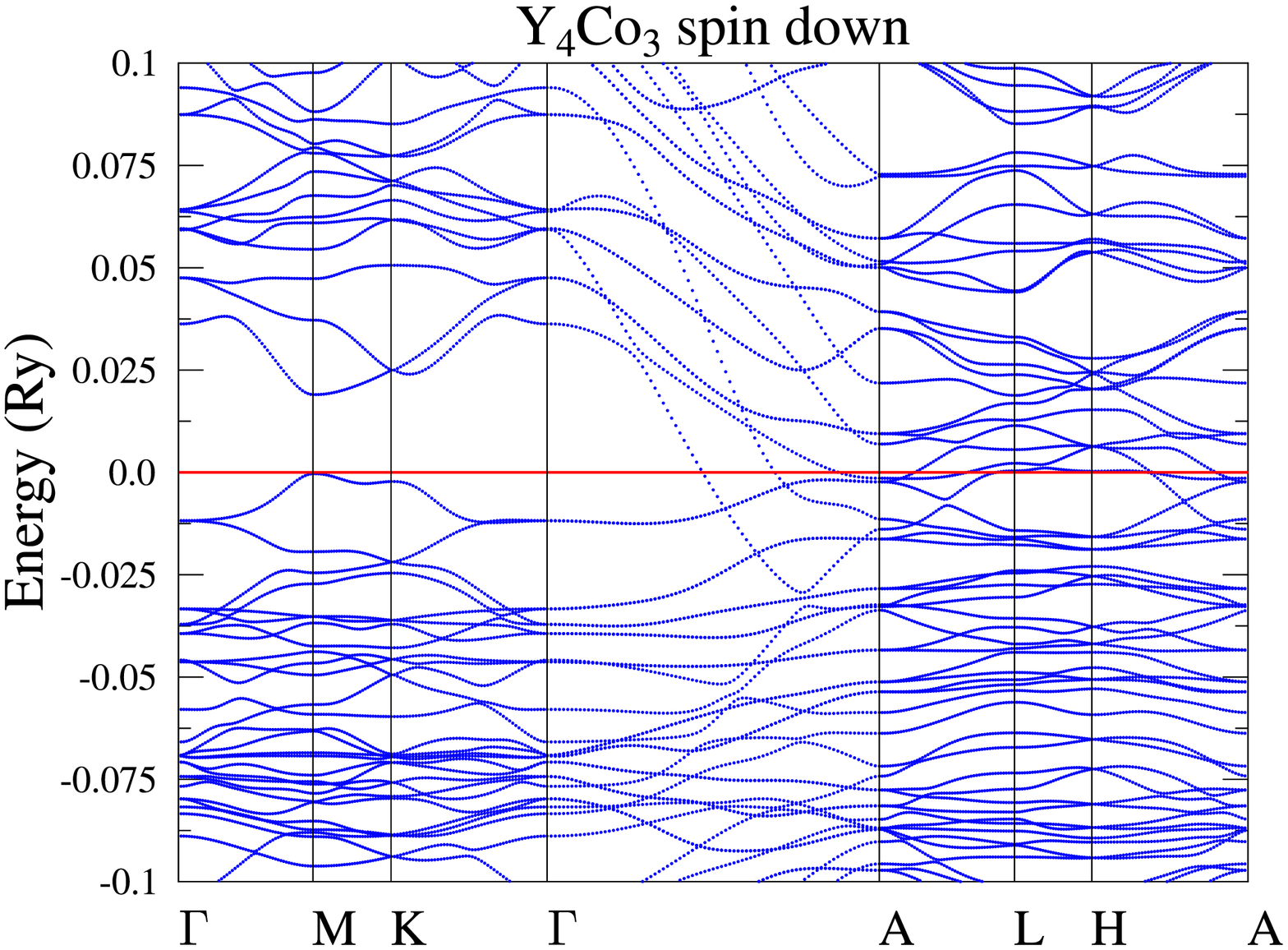}
\caption{\label{fig-bands}(Color online) Electronic dispersion curves, $E_F = 0$. Center: Brillouin zone with the high symmetry points.~\cite{bilbao,bilbao2}}
\end{figure*}

The specific character of the Y$_4$Co$_3$ electronic structure can be observed in the electron dispersion curves, shown in the
Fig.~\ref{fig-bands}. In the $\Gamma-M-K-\Gamma$ triangle no band crosses the Fermi level for both spin 
directions, and this behavior holds for any direction in the $k_z = 0$ plane, as was checked by extensive calculations 
along randomly chosen directions. Similar result was earlier observed from non spin-polarized 
calculations.~\cite{yco-jeong} The ferromagnetic state leads mainly to the slight shift of dispersion curves and  
some changes in details. Since, the  Y$_4$Co$_3$ system is predicted to possess an energy band gap in the $k_z = 0$ 
plane, it would be interesting to verify it (likely detectable on the ARUPS spectra, but due to lack of single crystals there are no such results available). The vanishing Fermi surface in the $k_z = 0$ plane is presumably responsible for the 
DOS valley formed around $E_F$ (Fig.~\ref{fig-tdos2}). In the $\Gamma-A$ direction (along the $k_z$ axis), 
almost linear and strongly dispersive bands cross $E_F$, while in the $A-L-H-A$ triangle, which is just a shift of 
the $\Gamma-M-K-\Gamma$ triangle to the Brillouin zone top, bands crossing $E_F$ become flat, which mostly gives 
rise to metallic-like properties of Y$_4$Co$_3$.

On the whole, the important result from presented spin-polarized KKR calculations, concerns the 
fact that there is no need to search for unconventional mechanism to explain magnetism in Y$_4$Co$_3$ 
system. The energetically unfavorable DOS peak just below $E_F$ (Fig.~\ref{yco-co3-dos}) on the 
single Co(3) atom, seems to be the reason of turning the system into the ferromagnetic state. 
In magnetic state, the spin-up peak is not seen near $E_F$, while in the spin-down DOS the peak is 
smaller and is expelled above $E_F$. In Ref.~\onlinecite{yco-shimizu}, devoted 
to the electronic structure and magnetism of Y$_9$Co$_7$, authors suggested two models of magnetism of this 
system, called A and B. Briefly, in the model A magnetism was attributed to the excess Co atoms on 
Y(1) and Y(2) sites, while in the model B the magnetism came from itinerant electrons of Co on (2b) 
sites. Authors~\cite{yco-shimizu} did not definitely conclude, which model was valid, but the model A was said to 
be the most probable, and the 'perfect Y$_9$Co$_7$ crystal' was expected to be 
paramagnetic. From our results we see, that itinerant ferromagnetism is an 
intrinsic property of Y$_4$Co$_3$ system. Hence, additional defects as the excess Co atoms on Y sites 
are not necessary to give rise ferromagnetism, even if instead of long range order in real material 
short range order appears, as suggested e.g. in Ref.~\onlinecite{yco-sro}. 
Similar behavior is expected to be valid for Y$_9$Co$_7$, since as we mentioned in Introduction, 
experimentally the magnetic properties of the Y-Co system do not change much between Y$_4$Co$_3$ and Y$_9$Co$_7$.~\cite{yco-liet}

To look closer on the magnetism of the Co(3) atom we can calculate the spin dependence of $d$ orbital 
occupation (actually, the diagonal elements of occupation matrix calculated using imaginary part of 
Green's function, defined as $\langle l_1m_1 | -{1 \over\pi} {\rm Im} G |l_2m_2\rangle$). 
The Co(3) (2b) site, being surrounded by Y(1) triangles in $z = \pm {1\over 4}$ planes, has a trigonal, $C_{3i}$ symmetry. 
Thus, there are two degenerated $E_g$ doublets, $\{d_{xy}, d_{x^2-y^2}\}$ and $\{d_{yz}, d_{xz}\}$ 
with occupation for both doublets approximately equal to 0.81e$_{\uparrow}$ and 0.76e$_{\downarrow}$.
The $d_{z^2}$ orbital ($A_g$ representation, $l=2, m=0$), aligned in 
the $z$ direction, is the highest-energy and least occupied orbital. It also has the highest level of 
polarization, with 0.79e$_{\uparrow}$ and 0.49e$_{\downarrow}$. Hence, about 60\% of magnetization 
comes from $d_{z^2}$ orbital and microscopically it is the main source of magnetism in this compound. 
The directional character of this orbital favors the simple ferromagnetic ordering of magnetic moments 
along the $z$-axis.

\begin{figure*}[t]
\includegraphics[height=0.32\textwidth]{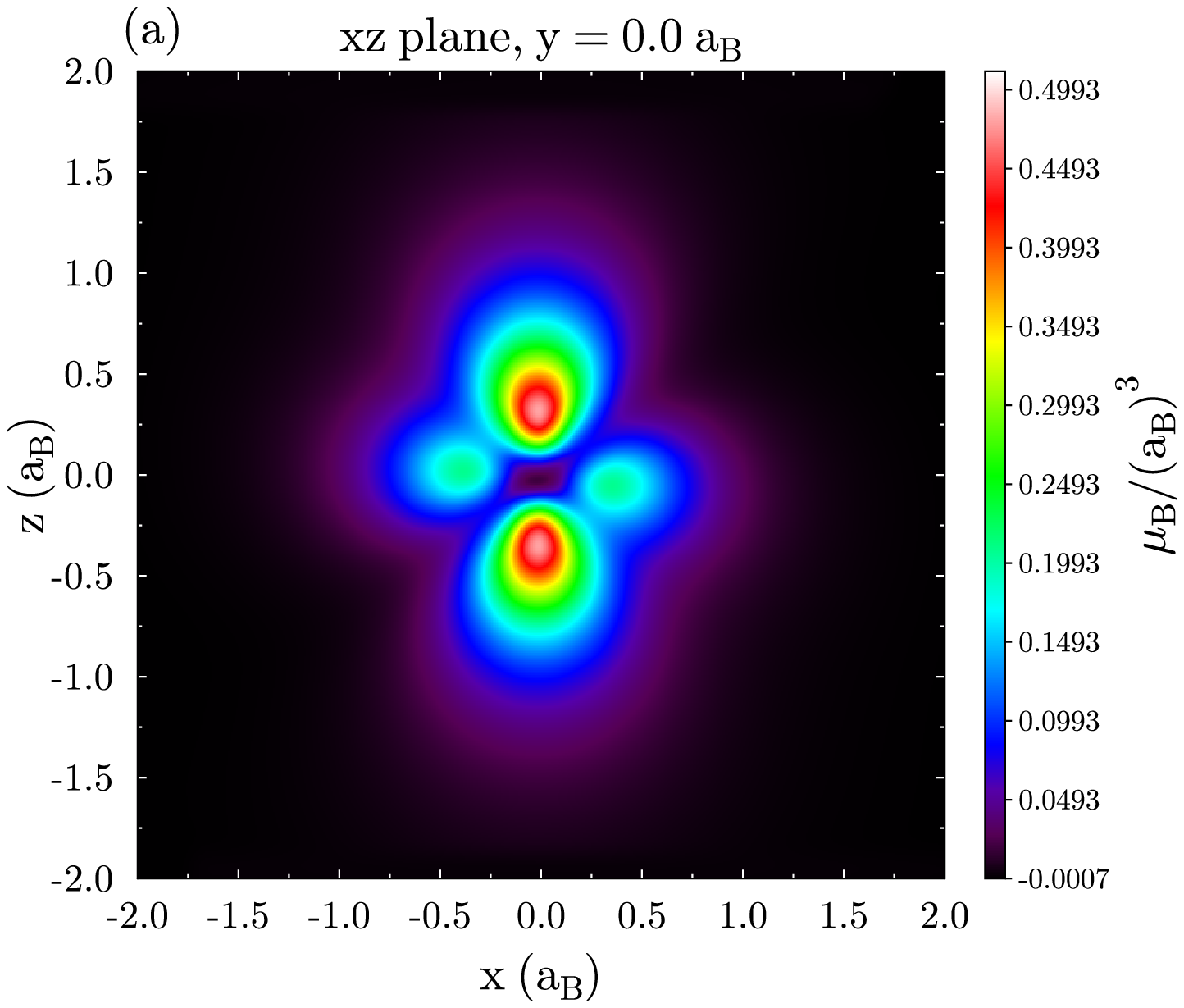}
\includegraphics[height=0.32\textwidth]{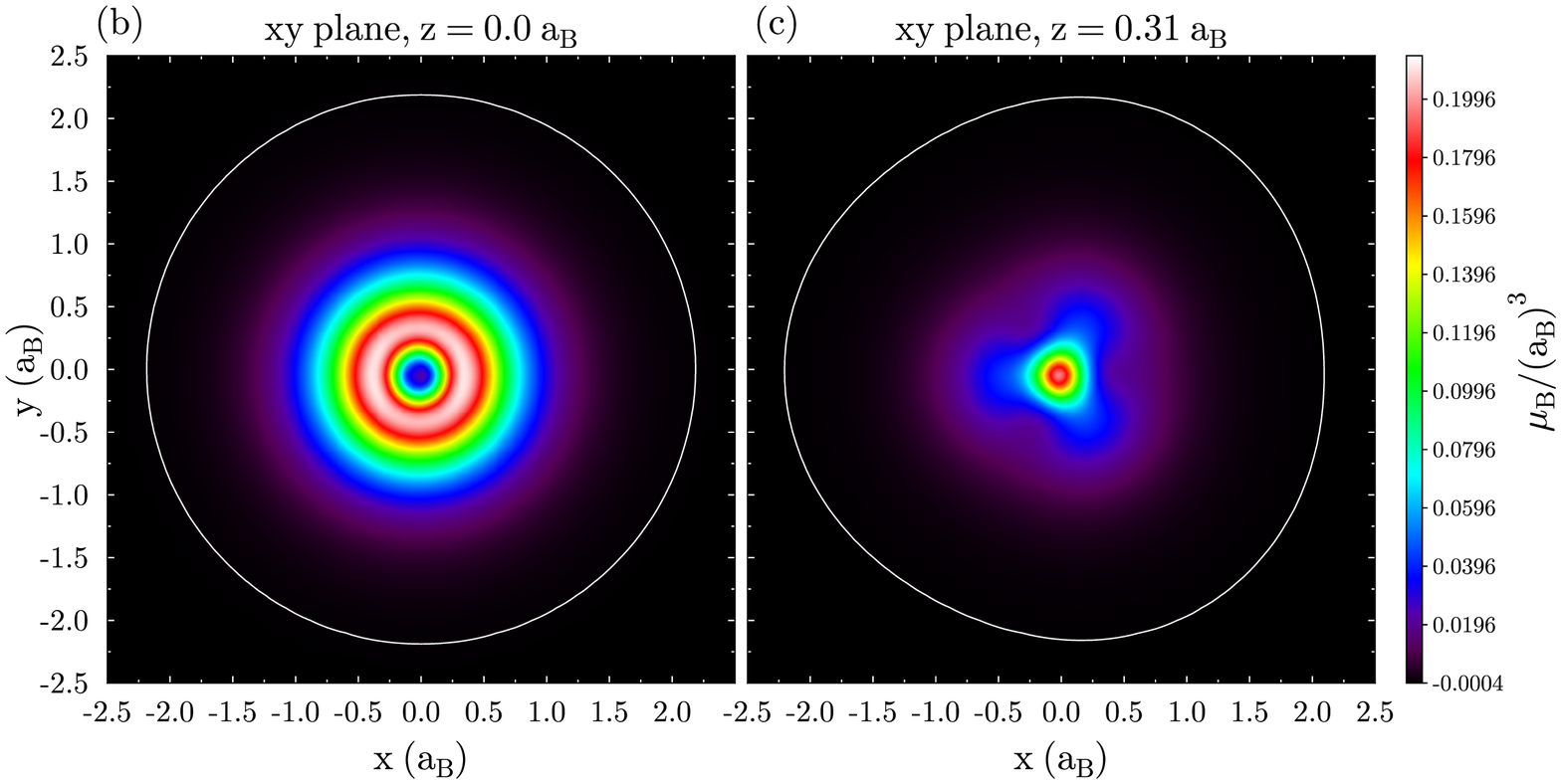}
\caption{\label{fig-charge}(Color online) Spin magnetization distribution in the (a) $y = 0$, (b) $z = 0$ and (c) $z = 0.04c$ (i.e. 0.31~$a_B$) planes.}
\end{figure*}

\begin{figure}[t]
\includegraphics[width=0.48\textwidth]{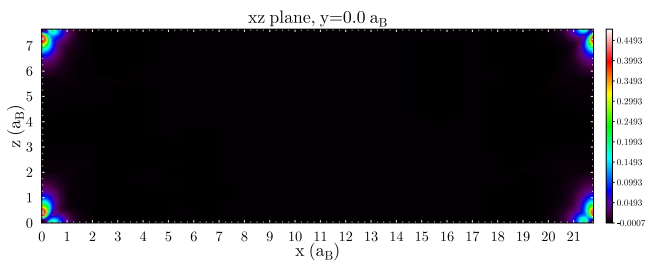}
\caption{\label{fig-charge2}(Color online) Spin magnetization distribution in the $y =0$ plane (face of the unit cell), showing that the charge density in the large areas of the unit cell is almost not polarized.}
\end{figure}

The fact that the magnetic moments are 'localized' only on Co(3) atoms can be well visualized on 
the spin magnetization contour maps, presented in Figs.~\ref{fig-charge} and~\ref{fig-charge2}. 
The spin magnetization is calculated as a difference between the spin-up and spin-down 
electron densities. The magnetic moments resides on the Co(3) corner atoms, 
while the rest of the presented $xz$ plane has magnetization close to zero (Fig.~\ref{fig-charge2}). 
The shape of magnetization around the Co(3) atom in the $xz$ plane (Fig.~\ref{fig-charge}a) confirms 
that the $3d_{z^2}$ orbital plays the major role in constitution of magnetic state. 
The $xy$ cross-sections well reflect the local symmetry of the magnetization, which should be 
trigonal ($C_{3i}$). Interestingly, in the $z = 0$ plane we may have an impression of cylindrical symmetry (Fig.~\ref{fig-charge}b), due to the equally distanced Y(1) triangles ($z = \pm {1\over 4}$), 
each rotated by 60$^o$. But, when we move off the $z=0$ plane, the triangle symmetry occurs
(Fig.~\ref{fig-charge}c).

Magnetism in Y$_4$Co$_3$ appears to be unusual. 
There is only one 'magnetic atom' in the complex 
unit cell, and twenty other atoms form diamagnetically polarized background.
The reason why other Co atoms are non-magnetic is understood from  
non spin-polarized partial DOS, where only Co(3) atom exhibits unstable DOS function due to 
the pronounced peak near $E_F$, while for Co(1) and Co(2) $E_F$ is located in the DOS minimum. 
Inducing magnetic moments on the Co(1) and Co(2) atoms would make the system energetically 
unfavorable.

Another peculiarity of this crystal structure is the arrangement of Co(3) atoms in chains, lying in 
channels formed by the octahedras of Y(1) atoms (see, Fig~\ref{fig:struct}).
Thus, the Co(3) atoms form a quasi-one-dimensional magnetic structure, a feature which was announced for this system~\cite{yco-press-apl,yco-sro} but not explored in the past. This low-dimensionality rises the question if the long-range ferromagnetic order can occur here, as the Mermin-Wagner theorem~\cite{mermin-wagner} predicts the lack of long-range magnetic order within a Heisenberg model with short-range interactions in one dimension. 
This of course does not completely rule out the presence of magnetic ordering for the three dimensional structures with quasi-one-dimensional chains, but the properties of such systems are affected by the low-dimensional character. Recently, the  quasi-one-dimensional ferromagnetism was investigated in some cobalt-based systems, like monoatomic Co chains on Pt substrate~\cite{co-1d-nature} or Co oxides, like BaCoO$_3$ or Ca$_3$Co$_2$O$_6$ (see, e.g. Refs.~\onlinecite{co-ox-1d, co-ox-1d2,co-ox-calc}). It is interesting to underline some structural similarities between Y$_4$Co$_3$ and aforementioned Co oxides. The Co(3) chains surrounded by Y(1) triangles, forming octahedras in $z$-direction, are similar to CoO$_6$ octahedras in oxides family. In both cases, the chains are running along $c$ axis of the hexagonal cell. The geometrical difference is that in Y$_4$Co$_3$ the intra- and inter-chain Co-Co distances are about twice as large as in the oxides. The large interchain distance in Y$_4$Co$_3$ (11.5~\AA) rather excludes the importance of interchain magnetic interaction, which is of great importance in oxides, and leads to two-dimensional antiferromagnetic effects.~\cite{co-ox-1d2}.
This quasi-one-dimensional character of magnetism in Y$_4$Co$_3$ makes this system even more interesting and opens new areas for future investigation.

\subsection{Ferromagnetism $vs.$ superconductivity}

The fact that most of the unit cell volume of Y$_4$Co$_3$ has negligible magnetization, allows to draw some qualitative 
conclusions on the possibility of coexistence of ferromagnetism and superconductivity. 
For example, in the $xz$ cross-section of the unit cell, presented in the Fig.~\ref{fig-charge2}, 
about 90\% of the area has magnetization lower than $10^{-3}$ $\mu_B/a_B^3$. 
For the whole unit cell, only 1.3~\% of the volume has magnetization higher than $10^{-3}$ $\mu_B/a_B^3$. 
Roughly speaking, if the electron density inside the unit cell has negligible polarization, in principle
it can not disturb forming singlet Cooper pairs. Certainly, the Cooper pairing acts in 
momentum space, so we should generally verify whether there are $({\bf k}_\uparrow, -{\bf k}_\downarrow)$ 
electrons near the Fermi level. Since the (${\bf k}, -{\bf k}$) degeneracy is ensured by the centrosymmetry of the 
unit cell, we should look for the $({\bf k}_\uparrow, {\bf k}_\downarrow)$ states. 
From the band structure plot e.g. in the $\Gamma-A$ direction we observe three bands crossing $E_F$. 
Two of these bands are almost the same for both spin-directions and they are crossing $E_F$ in points: 
${\bf k}_\uparrow = (0,0,0.2308) \frac{2\pi}{c}$, ${\bf k}_\downarrow = (0,0,0.2307) \frac{2\pi}{c}$ 
and ${\bf k}_\uparrow = (0,0,0.3422) \frac{2\pi}{c}$, ${\bf k}_\downarrow = (0,0,0.3410) \frac{2\pi}{c}$. 
These bands are mostly of the Co(1) and Y(2) character. So, one concludes that in principle there are 
electrons in the Y-Co trigonal prisms, which can form Cooper pairs and lead to singlet superconductivity. But 
more quantitative analysis of electron-phonon coupling is required to assess whether the 
BCS type superconductivity can appear in Y$_4$Co$_3$.

Nevertheless, our results and conclusions may support the model of coexistence of weak ferromagnetism 
and superconductivity in Y$_4$Co$_3$, with the ferromagnetism carried by chains of Co(3) atoms, 
screened by superconducting trigonal prisms of Y(1)-Co(1)-Y(2)-Co(2). 
Interestingly, this picture is qualitatively similar to the vortex lattice in type-II superconductors, since 
in both cases we have superconducting sample penetrated by the magnetic field lines, forming hexagonal 
lattice.

\subsection{Spin fluctuations\label{spinfl}}

The calculated value of Y$_4$Co$_3$ magnetization (0.13~$\mu_B$/(f.u.)), if compared to the measured $M = 0.045$~$\mu_B$/(f.u.) is 
about 2.5 times overestimated. Similarly, comparing the calculated Co(3) magnetic moment 
$\mu_{Co_3} = 0.55$~$\mu_B$ with the NMR estimation~\cite{yco-nmr1}  $\mu_{Co_3} = 0.23$~$\mu_B$ we 
get 2.5 times overestimation. 
This shows, that Y$_4$Co$_3$ can be a rare example of weak ferromagnetic system, where LDA tends 
to overestimate the tendency to magnetism, and suggests that it may be near the ferromagnetic quantum critical point.~\cite{weak_fm-mazin} The well-known similar examples are: ZrZn$_2$
Sc$_3$In and Ni$_3$Al/Ni$_3$Ga (see, e.g. Refs.~\onlinecite{zrzn2-mazin,sc3in-singh, sc3xb,ni3al-singh}), where due to strong spin 
fluctuations appearing in real samples, measured magnetic moments are much smaller than LDA (or GGA) 
values. Thus, our KKR-LDA results strongly support the classification of Y$_4$Co$_3$ as a weak 
itinerant ferromagnet with spin fluctuation effects, as already suggested, basing on the analysis 
of experimental results.~\cite{yco-spalek}
In Y$_4$Co$_3$, overestimation of magnetic moment is a little smaller, comparing to aforementioned cases, 
since in hexagonal Sc$_3$In LDA magnetic moment $\mu_{Sc} \simeq 0.35$~$\mu_{B}$ is about 7 times higher 
than experimental one,~\cite{sc3in-singh,phd} while the factor about 3 is established in 
ZrZn$_2$ (Ref.~\onlinecite{zrzn2-mazin}) and Ni$_3$Al (Ref.~\onlinecite{ni3al-singh}). This suggests that the spin fluctuations 
in Y$_4$Co$_3$ should be comparably weaker. The spin fluctuations strength parameter $\lambda\mathsf{_{sf}}$, which can be treated as the analog of the electron-phonon coupling constant $\lambda\mathsf{_{ph}}$, can be extracted from the electronic specific heat coefficient 
$\gamma$. The experimental value lies in the range of \makebox{$\gamma \simeq 3.1 - 3.4$~mJ/(mol at. K$^2$).
~\cite{yco-tarnawski,yco-heat2,yco-heat1}} Using the KKR-LDA value of total DOS at Fermi level 
(spin-polarized case), $N(E_F) = 229$~Ry$^{-1}$ we get $\gamma\mathsf{_{LDA}} = 1.89$~(mJ/mol at. K$^2$). 
Assuming that the band value is renormalized by the electron-phonon as well as the electron-paramagnon 
interaction (spin fluctuations), the formula $\gamma = \gamma\mathsf{_{LDA}}(1+\lambda\mathsf{_{ph}} + \lambda\mathsf{_{sf}})$, 
gives $\lambda\mathsf{_{ph}} + \lambda\mathsf{_{sf}} = 0.6 - 0.8$. This allows to safely put the upper limit 
$\lambda\mathsf{_{ph}} + \lambda\mathsf{_{sf}} < 1$. 
It is possible to have independent estimation of $\lambda\mathsf{_{ph}}$ and $\lambda\mathsf{_{sf}}$ if one 
admits that superconductivity is driven by phonons and the McMillan formula~\cite{mcm} can be applied here. 
We use the experimental value of $T_s$ and the renormalized McMillan formula to take into account the spin 
fluctuations:~\cite{carbotte} (see, also Ref.~\onlinecite{mosb-bw} and references therein)
\begin{equation}\label{eq:tc}
T_c =  \frac{\Theta_D}{1.45}\,\exp\left\{
-\frac{1.04(1+\lambda\mathsf{_{eff}})}
{\lambda\mathsf{_{eff}}-\mu\mathsf{_{eff}}^{\star}
(1+0.62\lambda\mathsf{_{eff}})}\right\},
\end{equation}
with the renormalization:
\begin{eqnarray}\label{eq:renorm}
\lambda\mathsf{_{eff}} &=& \lambda\mathsf{_{ph}}/(1+\lambda\mathsf{_{sf}}) \nonumber \\
\mu\mathsf{_{eff}}^{\star} &=& (\mu^{\star} + \lambda\mathsf{_{sf}})/(1+\lambda\mathsf{_{sf}}).
\end{eqnarray}
The experimental Debye temperature is $\Theta_D \simeq 215$~K.~\cite{yco-tarnawski} Noteworthy, the strong 
renormalization of Coulomb pseudopotential parameter $\mu^{\star}$ restricts the possible range of 
$\lambda\mathsf{_{sf}}$, e.g. we can get $T_s \simeq 2.5$~K for $\lambda\mathsf{_{sf}} = 0.1, 
\lambda\mathsf{_{ph}} = 0.7$ and $\mu^{\star} = 0.08$ (yielding 
$\mu\mathsf{_{eff}}^{\star} = 0.164$ and $\lambda\mathsf{_{ph}} + \lambda\mathsf{_{sf}} = 0.8$) or 
for  $\lambda\mathsf{_{sf}} = 0.15, \lambda\mathsf{_{ph}} = 0.85$ 
($\mu\mathsf{_{eff}}^{\star} = 0.20$ and $\lambda\mathsf{_{ph}} + \lambda\mathsf{_{sf}} = 1.0$).
Thus, we can estimate the electron-paramagnon interaction parameter to be of the order of 
$\lambda\mathsf{_{sf}} \sim 0.1$, but rather not higher than~0.2. 
This qualitative discussion shows that the spin fluctuations are likely present in Y$_4$Co$_3$, but 
they are not as strong as e.g. in 
Sc$_3$In, where $\lambda\mathsf{_{sf}} > 1$ can be deduced from the specific heat measurements 
in the magnetic field.~\cite{sc3in-heat1,sc3in-heat2} 
Noteworthy, similar measurements, i.e. the influence of high magnetic field on the electronic specific heat, 
would be very helpful to study the presence and strength of spin fluctuations in Y$_4$Co$_3$.

\subsection{Effect of pressure}

\begin{figure}[t]
\includegraphics[width=0.45\textwidth]{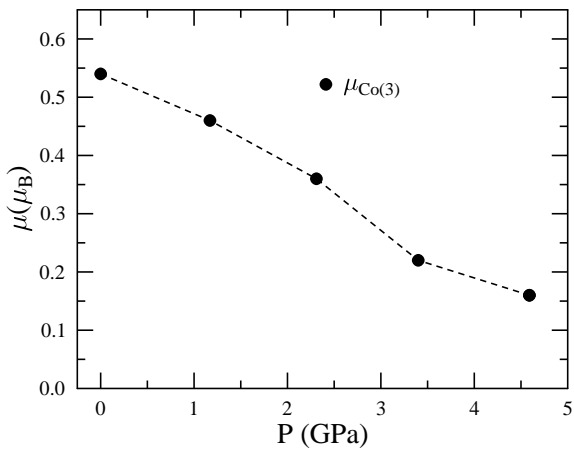}
\caption{\label{yco-mom} Evolution of the Co(3) magnetic moment with pressure.}
\end{figure}

\begin{figure*}[t]
\makebox{\includegraphics[width=0.30\textwidth]{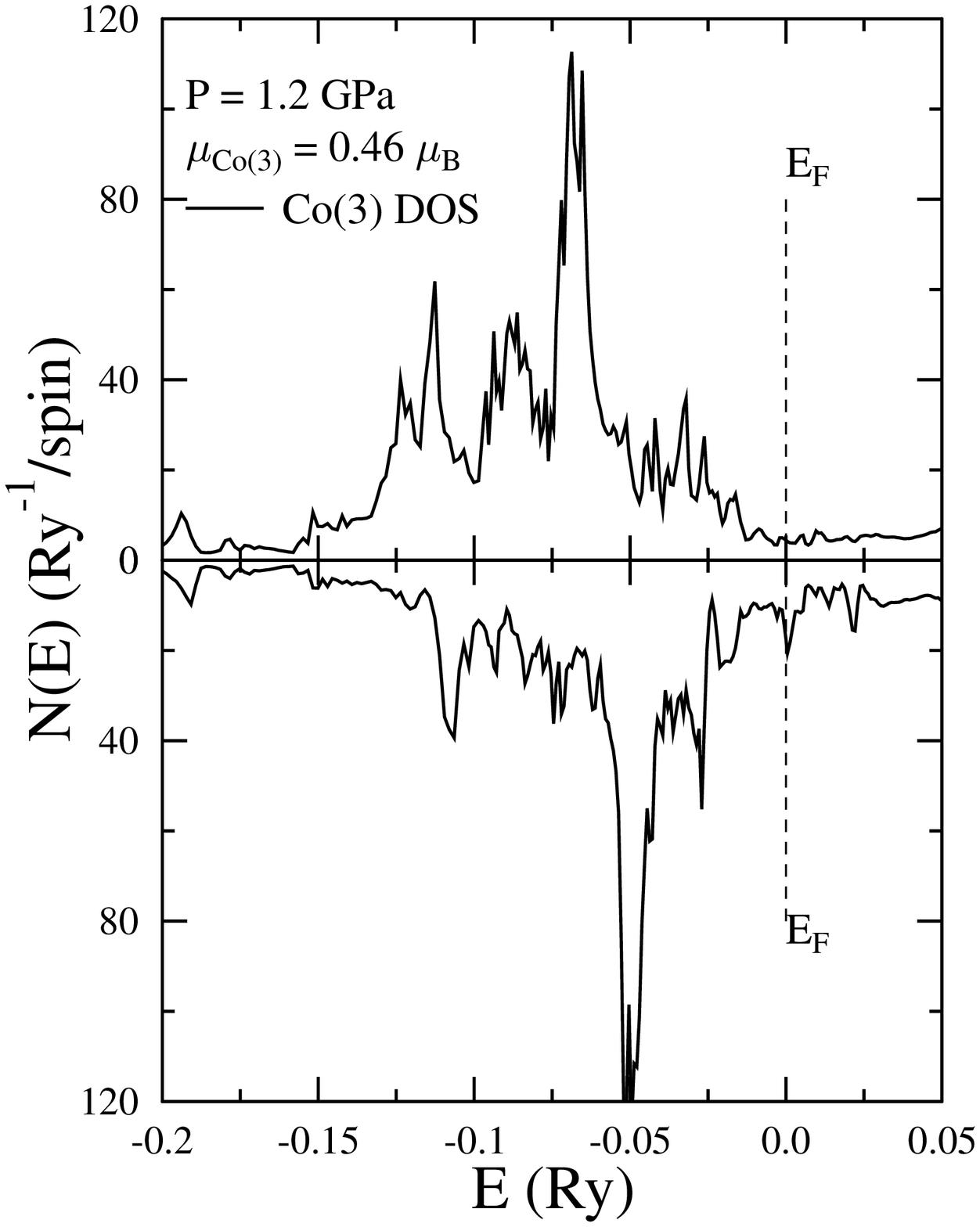} 
\includegraphics[width=0.30\textwidth]{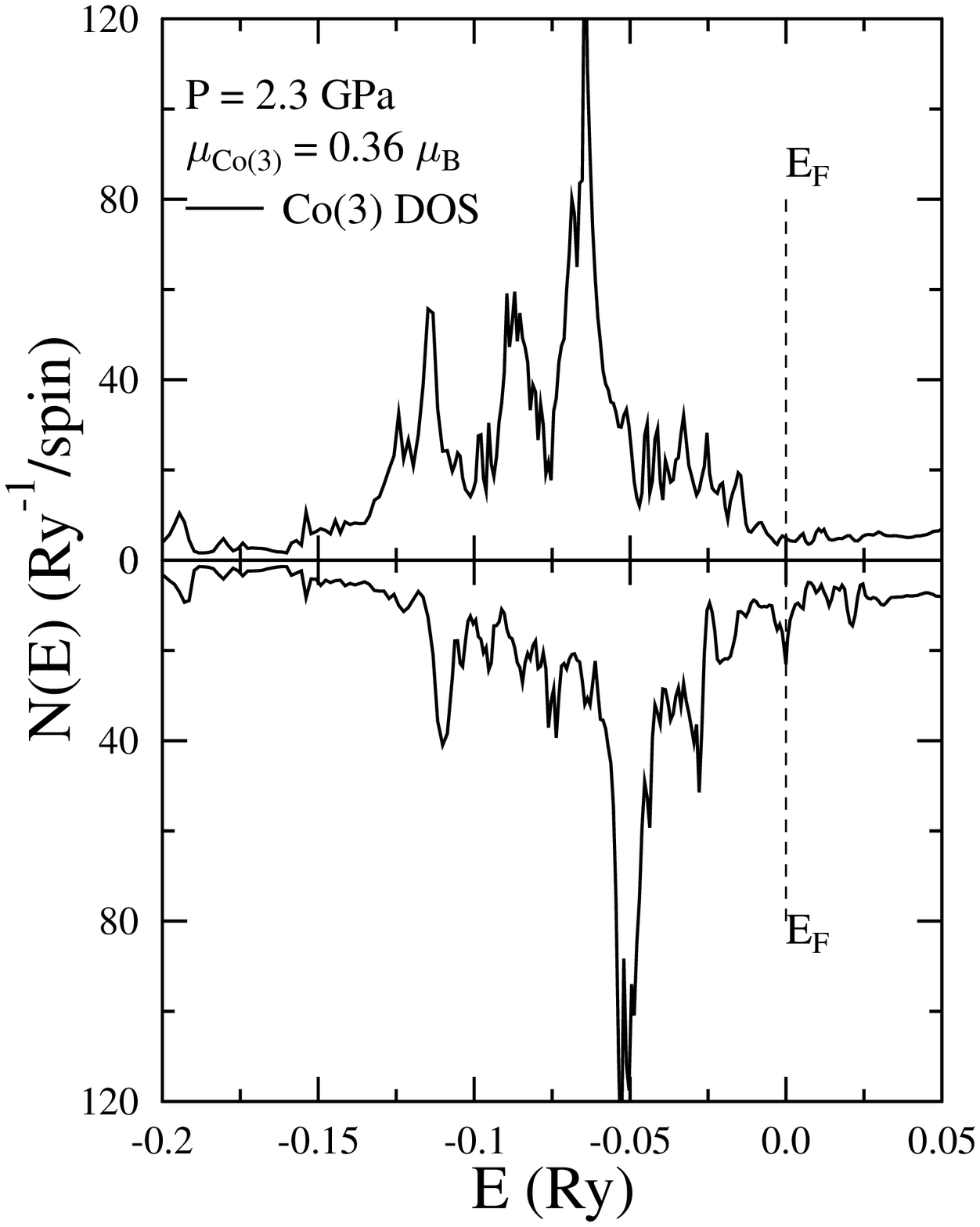} 
\includegraphics[width=0.30\textwidth]{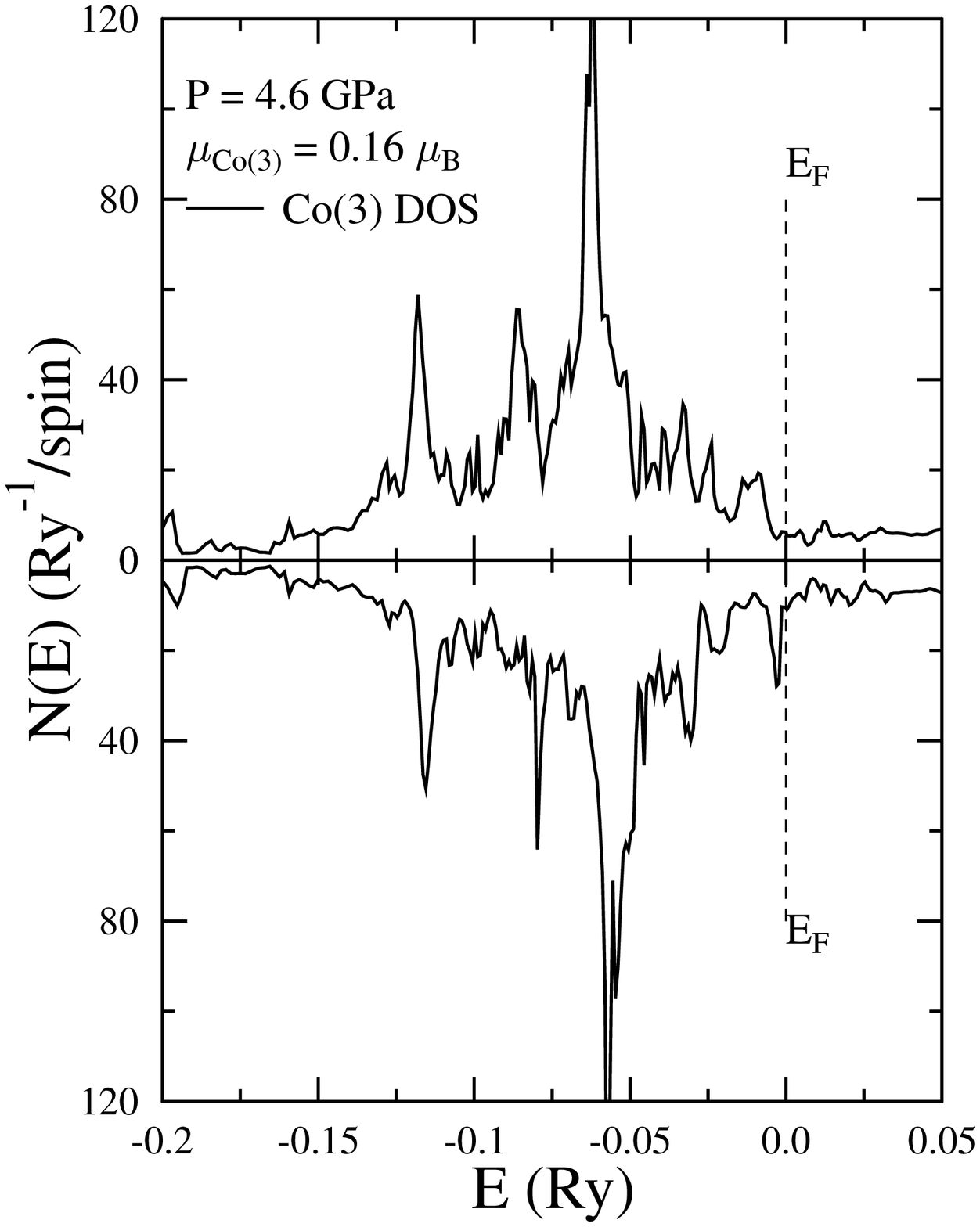}}
\caption{Evolution of the Co(3) DOS with pressure. Peak under $E_F$ is being formed.}\label{yco-co3-press}
\end{figure*}

The effect of pressure on the magnetic properties of Y$_4$Co$_3$ was also analyzed. We have performed 
calculations under 'quasi-hydrostatic' conditions. The $a$ and $c$ parameters were equally contracted 
in the range 0\%-2\% and all self-consistent calculations were started from the beginning. 
The bulk modulus, estimated under these conditions, is about $B = 75 \pm 5$~GPa, which is higher 
than the bulk modulus of yttrium ($\sim 40$ GPa~\cite{y-bulk}) and much smaller than the corresponding value 
for cobalt ($\sim 200$ GPa~\cite{co-bulk}). In calculations for the smaller unit cell volume, the magnetism is suppressed and the magnetic moment 
decreases rapidly, with the slope $\frac{\partial\mu}{\partial P} \simeq 0.08$~$\mu_B$/GPa, as seen in 
Fig.~\ref{yco-mom}. The reason for such behavior can be apparently seen from the evolution of the 
partial Co(3) DOS under pressure (Fig.~\ref{yco-co3-press}). When the unit cell volume decrease, 
the spin-up DOS peak under $E_F$ tends to constitute, while the spin-down 
peak, expelled above $E_F$ in magnetic state, tends to move below $E_F$. 
Thus, the pressure lowers the 
DOS polarization and the magnetic moment, which additionally confirms that the Co(3)-DOS peak is responsible for appearance
of magnetic moment on cobalt atoms on (2b) sites.

The suppression of magnetism under pressure from KKR calculations remains in agreement with the 
experimental trends observed in Y$_9$Co$_7$ (e.g. lowering of Curie temperature and 
magnetization,~\cite{yco-press,yco-press-apl} for Y$_4$Co$_3$ there are no such results available in the literature). Extrapolation of the curve from Fig.~\ref{yco-mom} 
to zero magnetic moment gives the critical pressure, where magnetism completely disappears, as
$p_c \simeq 7$ GPa. This value can be compared with very recent measurements of Y$_9$Co$_7$ 
under pressure~\cite{klimczuk-conf}, where the critical pressure was estimated to be $p_c \simeq 3$ GPa. 
The overestimation of the critical pressure is in line with the spin fluctuations and weak 
ferromagnetism in this system, as revealed from our calculations. 
Interestingly, if the zero-pressure value of the magnetic moment the NMR experimental 
estimation $\mu({Co_3}) \sim 0.23$~$\mu_B$ is accounted for, the theoretical slope 
$\frac{\partial\mu}{\partial P} \simeq 0.08$~$\mu_B$/GPa also predicts decrease 
of $\mu$ to zero for $p \simeq 3$~GPa.

\section{Summary and conclusions}

The results of the FP-KKR electronic structure calculations for Y$_4$Co$_3$ system were presented. 
The ferromagnetic state obtained from spin-polarized computations can be attributed to the single 
Co atom located on the (2b) site, being the only magnetic atom among 21 ones in the unit 
cell, and forming the quasi-one-dimensional magnetic chains.
The LDA values of the magnetization ($M \simeq 0.13 \mu_B$/f.u.), Co(3) magnetic 
moment ($\mu \simeq 0.55 \mu_B$) and critical pressure ($p_c \simeq 7$~GPa) are overestimated 
comparing with experiments, which can be tentativelly explained in terms of 
weak ferromagnetism with moderate spin fluctuations ($\lambda\mathsf{_{sf}} \sim 0.1$). 
The calculated spin magnetization distribution as well as other band structure parameters, tend to the
conclusion that the conventional, singlet-like superconductivity may coexist with ferromagnetism 
in Y$_4$Co$_3$, due to relatively weak magnetic moments arranged along thin chains 
(the unit cell edges) on the one hand, and the presence of non-polarized electrons at 
the Fermi level (filling most of the unit cell) on the other hand.
On the whole, the FP-KKR results confirmed that the quasi-one-dimensional magnetism is an intrinsic property 
of Y$_4$Co$_3$ and its coexistence with singlet superconductivity may be possible.

\begin{acknowledgments}
This work was partly supported by the COST P19 action and the Polish Ministry of Science and Higher 
Education (44/N-COST/2007/0).
\end{acknowledgments}

\bibliography{biblio}

\end{document}